\documentclass[a4paper,twocolumn,11pt,accepted=2023-01-18]{quantumarticle}
\pdfoutput=1
\usepackage[utf8]{inputenc}
\usepackage[english]{babel}
\usepackage[T1]{fontenc}
\usepackage{amsmath, amssymb}

\usepackage{tikz}
\usepackage{lipsum}
\usepackage{graphicx}
\usepackage{xcolor}
\usepackage{comment}
\usepackage{dsfont}
\usepackage{booktabs}
\usepackage{siunitx}

\usepackage{microtype}
\newcommand{\bra}[1]{\langle #1 \rvert}
\newcommand{\ket}[1]{\lvert #1 \rangle}
\newcommand{\be}{\begin{equation}}
\newcommand{\ee}{\end{equation}}
\newcommand{\beN}{\begin{equation*}}
\newcommand{\eeN}{\end{equation*}}
\newcommand{\ba}{\begin{align}}
\newcommand{\ea}{\end{align}}
\newcommand{\id}{\mathds{1}}

\DeclareMathOperator{\tr}{Tr}
\newcommand{\ignore}[1]{}

\usepackage[normalem]{ulem}

\usepackage{hyperref}
\hypersetup{%
	colorlinks=true, linktocpage=true, pdfstartpage=1, pdfstartview=FitH, pdfborder={0 0 0},%
	breaklinks=true, pdfpagemode=UseNone, pageanchor=true, pdfpagemode=UseOutlines,%
	plainpages=false, bookmarksnumbered, bookmarksopen=true, bookmarksopenlevel=1,%
	hypertexnames=true, pdfhighlight=/O,
	urlcolor=blue, linkcolor=blue, citecolor=blue
}

\usepackage[numbers,sort&compress]{natbib}

\begin{document}

\title{High-accuracy Hamiltonian learning via delocalized quantum state evolutions}

\author{Davide Rattacaso}
\email{davide.rattacaso@unina.it}
\affiliation{Dipartimento di Fisica ``E. Pancini'', Università di Napoli Federico II, Complesso di Monte Sant'Angelo, via Cinthia, Napoli 80126, Italy}
\orcid{0000-0001-8219-5806}
\author{Gianluca Passarelli}
\affiliation{CNR-SPIN, c/o Complesso di Monte Sant'Angelo, via Cinthia, Napoli 80126, Italy}
\orcid{0000-0002-3292-0034}
\author{Procolo Lucignano}
\affiliation{Dipartimento di Fisica ``E. Pancini'', Università di Napoli Federico II, Complesso di Monte Sant'Angelo, via Cinthia, Napoli 80126, Italy}
\orcid{0000-0003-2784-8485}

\maketitle

\begin{abstract}
Learning the unknown Hamiltonian governing the dynamics of a quantum many-body system is a challenging task.  In this manuscript, we propose a possible strategy based on repeated measurements on a single time-dependent state.
We prove that the accuracy of the learning process is maximized for states that are delocalized in the Hamiltonian eigenbasis. This implies that delocalization is a quantum resource for Hamiltonian learning, that can be exploited to select optimal initial states for learning algorithms. 
We investigate the error scaling of our reconstruction with respect to the number of measurements, 
and we provide examples of our learning algorithm on simulated quantum systems.
\end{abstract}

\section{Introduction}
Thanks to the enormous progress in manufacturing and controlling quantum devices made out of an increasingly large number of qubits, we are now entering the era of Noisy Intermediate-Scale Quantum technology~\cite{Preskill2018}. Relevant progresses have been made in controlling quantum degrees of freedom on different platforms~\cite{Ebadi2021,Chen2021,Scholl_2021}. However, to some extent the true Hamiltonian, governing the dynamics of these systems is often (at least) partially unknown. In this context, the major challenge is then to infer a realistic Hamiltonian model of the quantum system that can match the experimental data, guided by physical intuition. By querying the device (assumed a black box), one can measure the time evolution of several observables in order to learn the system Hamiltonian. This process, known as Hamiltonian learning, has been fundamental over the years for the validation of theoretical models, where the observation of the system evolution aims to characterize the unknown parameters of the model and establish its plausibility, and is now centralizing the attention of the scientific community due to its relevance for quantum technologies and quantum computation~\cite{Granade_2012,Wang_2015,Anshu2021,Burgarth_2011,Ruichao2017,h_learning_00,h_learning_01,h_learning_02,h_learning_0,h_learning_1,Bairey2019,Qi2019,Chertkov2018,Greiter2018,Hou2020,Cao2020,bisognano_1,bisognano_2,open_inverseprob_0,open_inverseprob_1,bairey2021,hangleiter2021,dutt2021active,Rattacaso2021}. Applications of Hamiltonian learning range from the verification of the performances of quantum devices~\cite{verification_0,verification_2,verification_3,verification_4,verification_5}, to quantum error correction~\cite{learning_for_correction}, to the design, characterization and calibration of quantum devices~\cite{learning_for_correction,hangleiter2021,dutt2021active}. 

\begin{figure}
    \centering
    \includegraphics[width=\columnwidth]{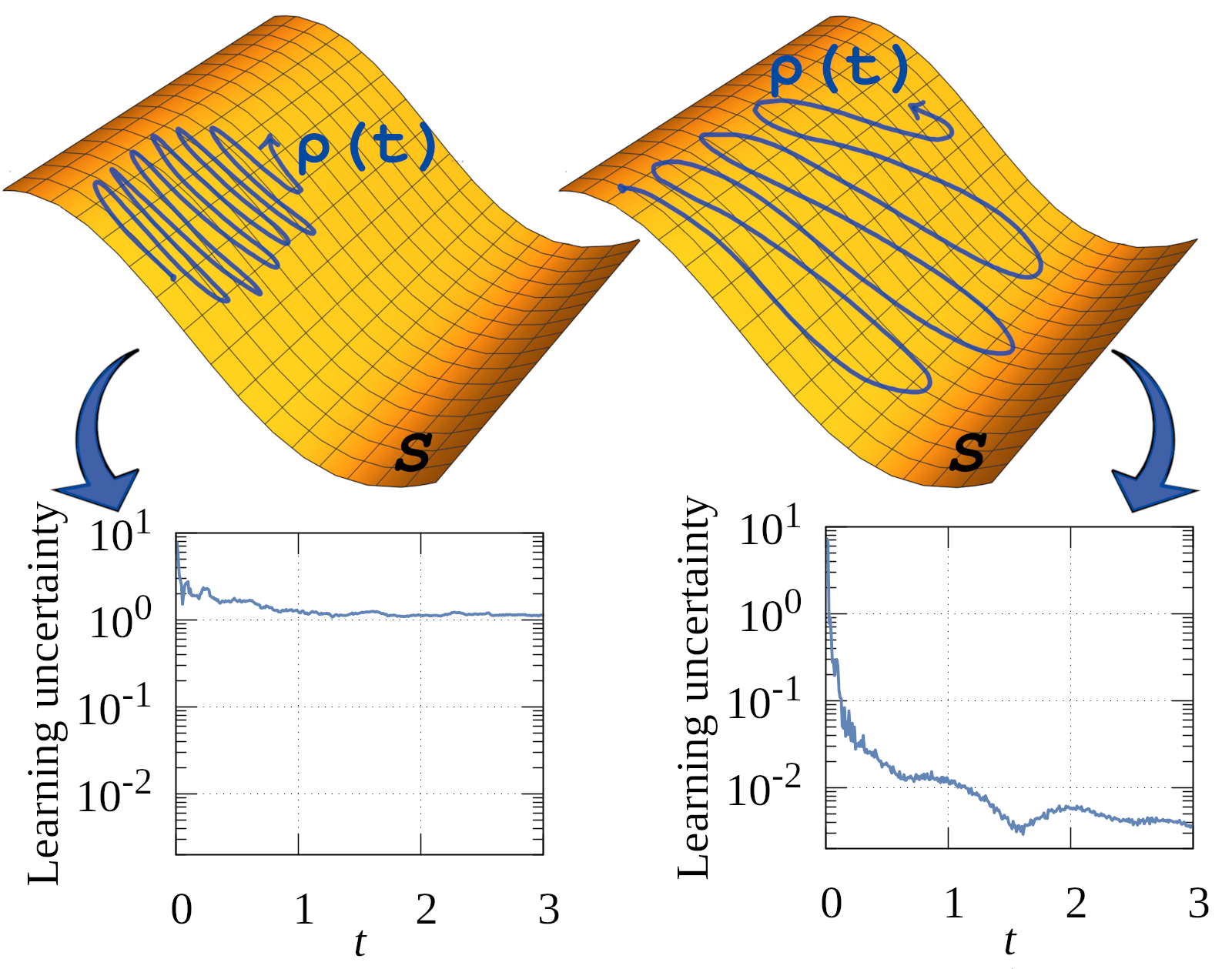}
    \caption{The larger is the capability of the state $\rho(t)$ to explore the space of states $S$, the larger is the amount of information obtained through the learning process, and, consequently, the smaller is the uncertainty in reconstructing the Hamiltonian.}
    \label{fig:main}
\end{figure}

The performance of a Hamiltonian learning algorithm is determined by the scaling of the relative uncertainty of the reconstructed Hamiltonian as a function of the computational effort required, which, in many relevant situations, is given by the number $N_S$ of experiments, or shots. Each shot begins with a state preparation and ends with a measurement in the computational basis. To fully reconstruct a quantum Hamiltonian, its action on a basis of the Hilbert space must be known. As a result, a number of initial states that is exponential in the systems size is needed for a full process tomography, leading to an exponential overhead in the number of shots required for learning.

When the Hamiltonian is the span of local interactions, it has been proven that a polynomial number of experiments is sufficient to fully reconstruct its couplings~\cite{poulin}. Significant examples in this direction are the reconstruction of the system Hamiltonian from short-time evolutions~\cite{bairey2021} and taking advantage from the exploitation of symmetries of the unknown Hamiltonian~\cite{hangleiter2021}. In these works, locality is a resource for the learning process.

In this paper, we focus on properties of the state (rather than the Hamiltonian) and we propose a novel algorithm that requires measurements of a single time-dependent quantum state. The main idea is to take benefit of quantum superposition. 
As depicted in Figure~\ref{fig:main}, indeed, some states are able to explore a significant portion of the space of states $S$ during their time evolution and therefore encode enough information to completely reconstruct the Hamiltonian generating their dynamics. Using analytical arguments and numerical examples, we  show that the accuracy of the method proposed is related to the delocalization of the initial state in the Hamiltonian eigenstates. To this aim, we prove an analytical relationship between the inverse participation ratio (IPR)~\cite{GogolinEisert,IPR2,IPR3} of the initial state, and the information matrix that measures the amount of information acquired in the learning process. Equally weighted superpositions of the Hamiltonian eigenstates explore a large sample of the space of states and provide the maximum amount of information about the system Hamiltonian. In other words, delocalization is a resource for Hamiltonian learning. This opens a new perspective on the application of quantum information theory to the study of out-of-equilibrium quantum systems~\cite{GogolinEisert,nonequilibrium}. 
As a proof of concepts, we apply our method to learn the Hamiltonian of systems of few superconducting qubits, highlighting its relevance to  gate-based quantum computation.  In this setting, we exploit state tomography to define a simple algorithm that clarifies the relationship between delocalization and learning. Full state tomography requires an exponential amount of resources in the system size. However, scaling to large system sizes is out of the purposes of this manuscript.

\section{Hamiltonian learning algorithm}\label{sec2}

To fully represent the information content of the system state, we define a basis $\mathcal{B}=\{O_\alpha\}$ for the space of Hermitian operators, orthonormal with respect to the Hilbert-Schmidt product $(A,B)=\tr(AB)$.  In this basis, the system density matrix $\rho(t)$ can be expanded as $\rho(t)=\sum_\alpha r_a(t)O_\alpha$ where $r_\alpha(t)\equiv\tr( O_\alpha\rho(t))$ are the components of $\rho(t)$ over $\mathcal{B}$ and are the expectation values of the observables $O_\alpha$ over the state $\rho(t)$. If we measured the coefficients $r_\alpha(t_n)$ at a collection of $N_T$ times $\{t_n\}\equiv\{0,\delta t,\dots,(N_T-1) \delta t\}$, we would have a time-dependent state tomography. Repeating  each measurement  $N_M$ times, the uncertainty on each observables  is $\sigma(r_\alpha(t_n))=\sqrt{\tr(O_\alpha^2\rho(t_n))/N_M}$. In superconducting quantum processing units, measurements are always performed in the computational basis, i.\,e., in the basis of simultaneous eigenstates of the single-qubit operators $\sigma_i^z$. Measuring observables that do not commute with $\sigma_i^z$, such as $\sigma_i^{x,y}$, can be done by first rotating the system state so that $\sigma_i^{x,y}\to \sigma_i^z$. This means that we need to perform $3$ measurements per qubit in order to have full information about the system state. As a consequence, in a system of $N_q$ qubits the number experiment shots needed for a full state tomography is $N_S=3^{N_q}N_TN_M$.

In general, the unknown Hamiltonian is a Hermitian operator that can be expanded in the (exponentially large in the system size) basis $\mathcal{B}$. However, realistic Hamiltonians often have symmetries and/or locality constraints that can be identified resorting to first-principle theoretical models~\cite{PhysRevA.101.052308}.  
Hence, we can write the unknown Hamiltonian as the span of a set of local Hermitian and traceless operators $\mathcal{B}_L=\{L_i\}$ that represent the relevant interactions between the constituents of the system: $H = \sum_i h_i L_i$. Remarkably, the locality constraint implies that the number of $L_i$ in $\mathcal {B}_L$ is at most polynomial in the system size ~\cite{Rattacaso2021}.

We want to exploit the information extracted from the state evolution via a full tomography to learn the couplings $ h_i$ of the system Hamiltonian $H $. Assuming that the system evolves unitarily according to the Liouville-von Neumann equation $\dot\rho = -i[H, \rho]$ (here and in the following $\hbar=1$), the system state at a time $t_{n+1}$ is related to $\rho(t_n)$ via the equation
\begin{equation}\label{01}
	\frac{\rho(t_{n+1})-\rho(t_n)}{\delta t}+i\sum_i h_i[L_i,\rho(t_n)]= R_n\delta t,
\end{equation}
where $R_n = \frac{-[ H,[ H,\rho(t^*)]]}{2}$ is the remainder of the Taylor expansion at the first order of $\rho(t_{n+1})$, for   $t^*\in[t_n,t_{n+1}]$.

In an ideal experiment, $\delta t$ could be made arbitrarily small and the uncertainty on the states $\rho(t_n)$ would vanish as well. As a consequence, the optimal couplings $ h_i^{\text{(opt)}}$ are the ones that minimize the Frobenius norm of the LHS of Eq.~\eqref{01} at each time, that is:
\begin{equation}
f(\vec h)=C-2\sum_ih_iB_i+\sum_{ij}V_{ij}h_ih_j,
\end{equation}
where
\begin{equation}
V_{ij} =-\sum_n \tr\left([L_i,\rho(t_n)][L_{j},\rho(t_n)]\right)\;,
\label{eq:tqcm0}
\end{equation}
$B_{j} = \sum_n \tr\left(-i[L_j,\rho(t_n)](\rho(t_{n+1})-\rho(t_n))/\delta t\right) $, and $C=\sum_n \tr\left((\rho(t_{n+1})-\rho(t_n))^2/\delta t^2\right)$. We call the matrix $V_{ij}$ the Total Quantum Covariance Matrix (TQCM).

The optimal couplings are such that the gradient of $f(\vec h)$ is null. When the TQCM is invertible these couplings can be written as
\begin{equation}\label{opt_coeffs}
	h_i^{\text{(opt)}} = \sum_{j} (V^{-1})_{ij}B_{j},
\end{equation}
Eq.~\eqref{opt_coeffs} can serve to our goal if the kernel of $V_{ij}$ is empty,  otherwise, the experimental data are insufficient to specify the system Hamiltonian, meaning that different Hamiltonians can produce the same observables. 

\section{Uncertainty estimation}

In real experiments, the expectation values $r_\alpha(t_n)$ are affected by statistical uncertainties. Moreover, the time interval $\delta t$ between measurements cannot be reduced to zero, and the remainder $R_n$ has to be considered as a systematic source of error for the estimated derivatives $\left(\rho(t_{n+1})-\rho(t_n)\right)/\delta t$. These contributions determine a total uncertainty $\delta B_i $ on the vector $B_i$. As a consequence, the Hamiltonian couplings are known up to an uncertainty $\delta h_i$ and any Hamiltonian $H = \sum_ih_iL_i$ with $\lvert h_i-h_i^\text{(opt)}\rvert <\delta h_i$ is compatible with the dynamics observed through the measurement of the local expectation values $r_\alpha(t_n)$.  Remarkably, since we perform tomography even at the initial time, state preparation errors do not directly affect the uncertainty on the reconstructed Hamiltonian.

As we show in Appendix \ref{sec:appendix}, both the statistical and the systematic part of the uncertainty $\delta B_i$ can be upper-bounded by a function that does not depend on the specific evolution of the system, but only on the Hamiltonian norm, the total number of repetitions $N_M$, time steps $N_T$ and the number of qubits $N_q$. Relating the uncertainty $\delta B_i$ to the uncertainty $\delta h_i$ through Eq.~\eqref{opt_coeffs} and considering the relationship between the vector norm of the exact couplings $\lVert\vec{ h}\rVert$ and the operator norm $\lVert  H\rVert_\text{op}$, we can extend this bound to the relative error
\be
\varepsilon\equiv\lVert \vec h^{\text{(opt)}}-\vec { h}\rVert /\lVert \vec { h}\rVert 
\ee
on the reconstructed Hamiltonian:
\begin{align}\label{eq:rel_err_scaling}
\varepsilon \leq &\sqrt{l\tr \left[\left(\frac{V_{ij}}{\lVert L\rVert_\text{op}^2 N_T}\right)^{-2}\right]}\nonumber\\ &\times \left(16 \frac{{(3/2)}^{\frac{N_q}{2}}}{\sqrt{N_S}} +4\lVert  H\rVert _\text{op}\delta t\right),
\end{align}
where $l$ is the number of couplings of the Hamiltonian and, without loss of generality, we suppose that $\lVert L_i\rVert_\text{op}=\lVert L\rVert_\text{op} \forall i$. Remarkably, this reconstruction error is not affected by the derivative uncertainty, inversely proportional to the time step, which decreases the accuracy of the learning protocols based on short-time evolution\cite{bairey2021}. This feature can represent a significant advantage in characterizing devices with a high temporal resolution.

 In Eq.~\eqref{eq:rel_err_scaling}, the upperbound on the relative error only depends on the system evolution through the eigenvalues of the TQCM. In particular, the larger are these eigenvalues, the larger is the amount of information about the system Hamiltonian that we gain by observing the state evolution. Conversely, when some eigenvalue of the TQCM goes to zero and the matrix is not invertible, the uncertainty diverges, signaling that the information acquired during the experiment is not sufficient for Hamiltonian learning. We can clearly see the statistical meaning of the TQCM: Eq.~\eqref{eq:rel_err_scaling} is analogous to a Cramer-Rao bound on the error on the estimated Hamiltonian~\cite{ElementsOfInformationTheory,dutt2021active}, where the TQCM takes the role of an information matrix. While the spectrum of $V_{ij}$  estimates the total amount of information about the Hamiltonian that we have gained by observing the state during its evolution, at any fixed time $t_n$, the spectrum of
\begin{equation}\label{eq:QCM}
V_{ij,n}\equiv-\tr\left([L_i,\rho(t_n)][L_{j},\rho(t_n)]\right)
\end{equation}
estimates the amount of information about the Hamiltonian gained by observing the short-time evolution around $t_n$. Both $V_{ij}$ and $V_{ij,n}$ depend on the initial state preparation and on the consequent evolution. The quantum state preparation of the  initial state is crucial to the success of the algorithm,  optimal initial states are those maximizing the amount of information gained by observing the evolution, hence, minimizing the uncertainty on the reconstructed Hamiltonian.

\section{Information and inverse participation ratio}

Schrodinger's evolution often delocalizes states in the Hilbert space. In this section, we show that the more the state samples the Hilbert space, the more information we gain about the system Hamiltonian.

As shown in Eq.~\eqref{eq:rel_err_scaling}, the largest contribution to the relative error comes from the minimum eigenvalue of the TQCM. The TQCM is the sum over the different time steps of the covariance matrices $V_{ij,n}$. Following the definition of Eq.~(\ref{eq:QCM}), these are positive semi-definite. Evolution of highly delocalized states generates very different covariance matrices at different times. In this case, we speculate that the eigenvalues of the TQCM exponentially increase for an initial transient, corresponding to the time spent by the state $\rho(t)$ in exploring the space of states before returning close to its previous orbit. The larger is the sample of the space of states explored by $\rho(t)$ during its evolution, the larger will be the amount of information on the system Hamiltonian gained by observing the evolution of $\rho(t)$.

Quantitatively, we show that a good estimation of the information obtained in the learning process is the IPR of the initial state in the Hamiltonian eigenstates~\cite{GogolinEisert,IPR2,IPR3}. If the system Hamiltonian is $ H=\sum_\alpha E_\alpha \ket{\alpha}\bra{\alpha}$ and the initial state of the system is $\ket{\psi}=\sum_\alpha a_\alpha\ket{\alpha}$, the IPR is defined as $\text{IPR}(\psi, H)=\sum_\alpha |a_\alpha|^4$. 
Therefore, the IPR measures the spreading of the initial state in the Hamiltonian eigenstates: the lower is the IPR, the more the initial state spreads out. This is also an estimation of the capability of the state of sampling the Hilbert space uniformly during the evolution, in analogy with the ergodic hypothesis. Indeed, the time average of an observable $A$ during the evolution is $
\bar A=\tr(\bar \rho A)$, where $\bar \rho =\sum |a_\alpha|^2\ket{\alpha}\bra{\alpha}$ is the dephased state. As a consequence, when the IPR is minimum ($\text{IPR}_\text{min} = 1/2^{N_q}$) all the populations of $\bar\rho$ are equal and the time average becomes an average on the energy eigenstates with equal weights.

The relationship between the uncertainty on the reconstructed Hamiltonian and the IPR will be clear in the numerics that follow. However, a simple general argument is given here, showing that the states with a small IPR are associated with large eigenvalues of the TQCM and, due to Eq.~\eqref{eq:rel_err_scaling}, with a small error of the reconstructed Hamiltonian.

Please note that the in order to calculate the IPR, one needs to know the Hamiltonian eigenstates $\ket \alpha$. This may sound odd, as the Hamiltonian itself is unknown in general. However, in our examples we apply our method to learn ``known'' Hamiltonians, and we can explicitly calculate the IPR.  In possible practical applications, this is not the case, but one could resort to adaptive learning approaches as in Ref. \cite{dutt2021active}. Starting from a state minimizing the IPR over the eigenstates of a guessed Hamiltonian, one can iteratively obtain better estimates of the target Hamiltonian.

We consider $\rho(t)$ as a pure state $\rho(t)=\ket{\psi(t)}\bra{\psi(t)}$, where $\ket{\psi(0)}=\sum_\alpha a_\alpha\ket{\alpha}$ and $\ket{\alpha}$ are the eigenstates of the unknown system Hamiltonian. The eigenvalues  $\omega_i$  of the TQCM  can be  written as $\omega_i=\sum_{jj'}V_{jj'}a^{(i)}_ja^{(i)}_{j'}$, in terms of the corresponding normalized eigenvectors $a^{(i)}_j$. Defining the local operators $A_i\equiv \sum_ja^{(i)}_jL_j$, the eigenvalues $\omega_i$ of the TQCM of Eq.~\eqref{eq:tqcm0} read:
\begin{align*}
\omega_i&=-\sum_n^{N_T} \tr\left([A_i,\rho(t_n)][A_i,\rho(t_n)]\right)\\
&=2 \sum_n^{N_T}\left(\langle \psi(t_n)| A_i^2|\psi (t_n)\rangle - \langle \psi(t_n)|A_i|\psi(t_n)\rangle^2\right).
\end{align*}

If the time-step $\delta t$ is sufficiently small, this sum can be approximated by an integral:
Considering that $\ket{\psi(t)}=\sum_\alpha a_\alpha\ket{\alpha}e^{-iE_\alpha t}$, for $\delta t\to0$ these eigenvalues can be written as integrals of local correlations:
\begin{align}
\omega_i &\approx
\frac{2 N_T}{T}\int_0^T dt\left(\langle\psi(t)|A_i^2|\psi_n\rangle-\langle\psi(t)|A_i|\psi(t)\rangle^2\right)\nonumber\\
&=\frac{2 N_T}{T}\int_0^T dt\Big(\sum_{\alpha \beta}a_\alpha a_\beta^*\langle \alpha|A_i^2|\beta\rangle e^{-it(E_\alpha-E_\beta)}\nonumber\\
&-\sum_{\alpha \beta \gamma \delta}a_\alpha a_\beta^*a_\gamma a_\delta^*\langle \alpha|A_i|\beta\rangle\langle\gamma|A_j|\delta\rangle\nonumber\\
&\times e^{-it(E_\alpha-E_\beta+E_\gamma-E_\delta)}\Big).
\end{align}

At this point we impose a non-degeneracy and non-resonance condition on the system Hamiltonian $ H$. The non-resonance condition consists in the fact that $E_\alpha-E_\beta+E_\gamma-E_\delta=0$ if and only if $E_\alpha=E_\beta$ and $E_\gamma=E_\delta$ or $E_\beta=E_\gamma$ and $E_\alpha-E_\delta$. This condition is automatically satisfied by any Hamiltonian that is not fine-tuned, so we can assume its validity for the real unknown Hamiltonian. In this way, after a time transient $T_e$ that is the equilibration time of the system~\cite{GogolinEisert}, the oscillating terms vanish and the previous equation becomes
\begin{align}
\omega_i &\approx
2N_T\Big[\sum_\alpha |a_\alpha|^2\langle \alpha|A_i^2|\alpha\rangle\nonumber\\
&-\sum_{\alpha\beta}|a_\alpha|^2|a_\beta|^2\Big(\langle\alpha|A_i|\alpha\rangle\langle\beta|A_i|\beta\rangle+\lvert\langle\alpha|A_i|\beta\rangle\rvert^2\Big)\Big].
\end{align}
or, equivalently,
\begin{equation}\label{eq:IPRvsTQCM}
\omega_i \approx
2N_T[\tr(\bar \rho A_i^2)-(\tr^2(\bar\rho A_i)+\tr(\bar\rho^2 A_i^2))].
\end{equation}

After the equilibration time these eigenvalues become linear in the number of time steps, with a coefficient $[\tr(\bar \rho A_i^2)-(\tr^2(\bar\rho A_i)+\tr(\bar\rho^2 A_i^2))]$ that corresponds to a measure of variance for $A_i$ in the dephased state $\bar\rho$. The positive contribution to this variance comes from the term $\tr(\bar \rho A_i^2)$, while the negative contributions come from $\tr^2(\bar\rho A_i)$ and $\tr(\bar\rho^2 A_i^2)$. When the IPR of the system state is near to its minimum, the dephased state is well approximated by the totally mixed state and, since the $L_i$'s are traceless, the term $\tr^2(\bar\rho A_i)$ vanishes. The remaining negative contribution also decreases in magnitude with the IPR, that can be written as $\tr(\bar\rho^2)$. We can conclude that minimizing the IPR is a good strategy to generate larger eigenvalues $\omega_i$ of the TQCM.

The optimal initial states, minimizing the IPR and therefore optimizing the learning process, are thus
\be
\ket{\psi_\text{opt}}=\sqrt{2^{-N_q}}\sum_\alpha e^{i\phi_\alpha}\ket{\alpha}
\ee
where the $\phi_\alpha$'s are arbitrary phases that we can fix to zero.

Due to the beneficial effect of delocalization, one could wonder if mixed states can also be considered a resource for learning algorithms. However, since the TQCM in Eq.~(\ref{eq:tqcm0}) is related to the trace of $\rho^2$, we expect states with smaller purity to have a smaller TQCM and to determine a worse accuracy.

\section{Simulations}
We have applied our Hamiltonian learning method to some few-qubit problems, in order to verify our predictions about the error scaling in Eq.~\eqref{eq:rel_err_scaling} and the relationship between the TQCM and the IPR in Eq.~\eqref{eq:IPRvsTQCM}. After focusing on a simple two-qubit problem, we discuss a three-qubit model with random couplings and we use Qiskit~\cite{Qiskit} to show how to apply our method to a real quantum processor.

In order to apply out learning procedure,  we  first choose a Hamiltonian $H$ and generate the evolution of the expectation values of the basis elements $\{O_\alpha\}$ by numerically integrating the Liouville equation for a set of initial states, corresponding to different initial configurations for the experiment with different IPRs. The effect of the statistical error is simulated by adding a uniform random noise with amplitude $1/\sqrt{2^{N_q} N_M}$ to each expectation value. Then, given the simulated expectation values $r_\alpha(t_n)$, we apply our method to find the optimal Hamiltonian $H^{\text{(opt)}} = \sum_i  h_i^{\text{(opt)}}L_i$. 
The success of the learning procedure can be checked comparing $H$ with $H^{\text{(opt)}}$. In particular  we analyze  the relative error $\varepsilon$ and the TQCM for each initial state and for different numbers of times steps $N_T$, corresponding to different total observation times. We  also estimate the relationship between the IPR and the information gained in the experiment, measured by the eigenvalues of the TQCM. Finally, we also calculate the optimal initial state $\ket{\psi_\text{opt}}$ for each Hamiltonian and exploit it for an optimal learning.

\subsection{Cross-resonance gate}

We focus on a quantum system representing a two-qubit device governed by the typical Hamiltonian of a cross-resonance gate~\cite{PhysRevB.81.134507,PhysRevLett.107.080502,PhysRevA.93.060302,PhysRevA.101.052308}. The implementation of this gate, consisting in two transmon qubits coupled by a bus
resonator, is fundamental to realize the CNOT gate for universal quantum computation. The Hamiltonian that we want to reconstruct is
\begin{align}\label{eq:crg_ham}
	H &= - 1.548 \, \id\otimes\sigma_x - 0.004 \, \id\otimes\sigma_y + 0.006\,  \id\otimes\sigma_z\nonumber\\ &\quad+ 9.578\,  \sigma_z\otimes\id + 5.316\,  \sigma_z\otimes\sigma_x \nonumber\\ &\quad - 0.225 \, \sigma_z\otimes\sigma_y - 0.340 \, \sigma_z\otimes\sigma_z,
\end{align}
where the couplings are taken from Ref.~\cite{tutorial} and the energies are expressed in $\si{MHz}$.
The optimal Hamiltonian is then found as the span of the operators $\{L_i\}=\{\id\otimes\sigma_x, \id\otimes\sigma_y, \id\otimes\sigma_x, \sigma_z\otimes\id, \sigma_z\otimes\sigma_x, \sigma_z\otimes\sigma_y, \sigma_z\otimes\sigma_z\}$. This choice is justified by first-principles studies~\cite{PhysRevA.93.060302}.

The Hamiltonian learning algorithm is performed with a time step $\delta t=0.01$, $N_M=1000$ measurement repetitions and a total number of shots $N_S=3\times10^6$, for four initial states with different IPR. The reconstructed Hamiltonian couplings for each initial state  (with its IPR) are shown in Table \ref{table:00}. We observe that when the initial state has a small IPR, the accuracy is maximized.

\begin{figure*}[t]
    \centering
    \includegraphics[width=2\columnwidth]{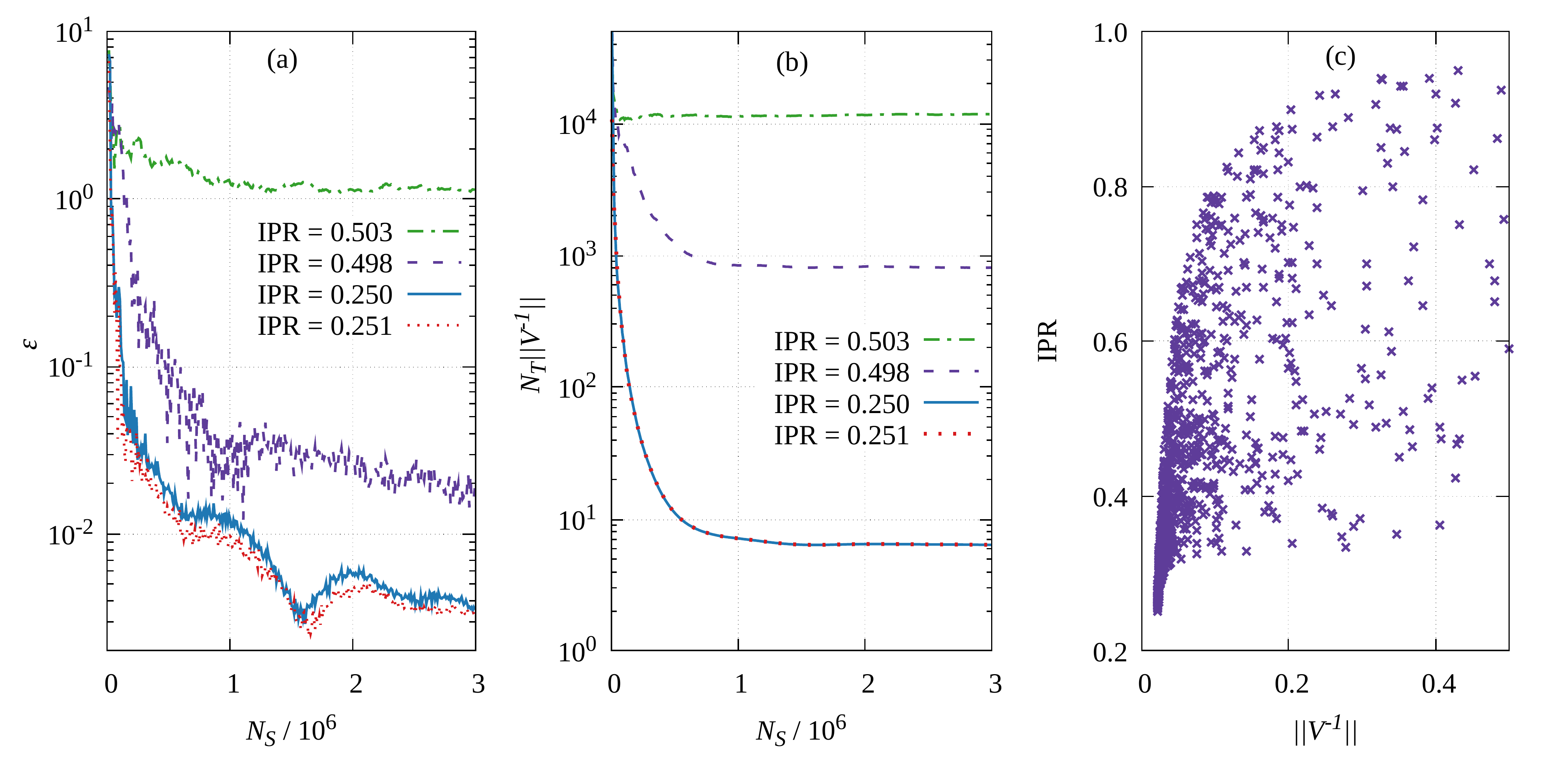}
    \caption{(a) relative reconstruction error and (b) Frobenius norm of the inverse TQCM multiplied by $N_T$, as a function of $N_S$ for the cross-resonance gate Hamiltonian in Eq.~(\ref{eq:crg_ham}), with $\delta t = 0.01$, $N_M=1000$ and $N_S=9\times N_M \times N_S$, for different initial states with different IPR. (c) IPR and Frobenius norm of the inverse TQCM for a collection of random states at large final time for the cross-resonance gate Hamiltonian, with $N_T=300$.}
    \label{fig:err_2q}
\end{figure*}

\begin{table}[b]
\centering
\caption{Estimated Hamiltonian coupling for different initial states with the corresponding IPR in parentheses, with $\delta t = 0.01$, $N_M=1000$ and $N_S=3\times10^6$. We define $\ket{\rightarrow\rightarrow}\equiv(\ket{\uparrow}+\ket{\downarrow})\otimes(\ket{\uparrow}+\ket{\downarrow})/2$ and $\ket{\psi_\text{Bell}^+}\equiv(\ket{\uparrow\uparrow}+\ket{\downarrow\downarrow})/\sqrt{2}$.}
\label{table:00}
\resizebox{\columnwidth}{!}{%
\begin{tabular}{cccccc}
	\toprule
	Term & Target & $\ket{\uparrow\uparrow}$ & $ \ket{\rightarrow\rightarrow}$ & $\ket{\psi_\text{Bell}^+}$ & $\ket{\psi_\text{opt}}$\\[1ex]
	 & & ($0.503$) & ($0.498$) & ($0.251$) & ($0.25$)\\
	\midrule
	$\id\otimes\sigma_{x}$ & $-1.548$ & $2.142$ & $-1.143$ & $-1.543$  & $-1.542$ \\
	$\id\otimes\sigma_{y}$ & $-0.004$ & $1.021$ & $-0.024$ & $-0.011$ & $0.000$ \\
	$\id\otimes\sigma_{z}$ & $0.006$ & $1.073$ & $-0.017$ & $-0.009$ & $0.023$\\
	$\sigma_{z}\otimes\id$ & $9.578$ & $-1.418$ & $9.748$ & $9.556$ & $9.553$\\
	$\sigma_{z}\otimes \sigma_{x}$ & $5.316$ & $1.627$ & $5.089$ & $5.301$ & $5.295$\\
	$\sigma_{z}\otimes \sigma_{y}$ & $-0.225$ & $-1.254$ & $-0.218$ & $-0.216$ & $-0.237$ \\
	$\sigma_{z}\otimes \sigma_{z}$ & $-0.340$ & $-1.409$ &  $-0.328$ & $-0.324$ & $-0.357$ \\
    \bottomrule
	\end{tabular}
}
\end{table}

In Fig.~\ref{fig:err_2q}(a) we show the behavior of the relative reconstruction error as a function of the number of experiment shots $N_S$. $N_S$ is increased by increasing the number of time steps $N_T$ and, consequently, the observation time. Different curves represent states with different IPR. The numerical simulations confirm our predictions: after a transient in which the error decreases exponentially in the number of time steps, the final error converges to values that are smaller for states with a smaller IPR. Optimal results are obtained when the initial state is $\ket{\psi_\text{opt}}$. In Fig.~\ref{fig:err_2q}(b), we show the behavior of the Frobenius norm of the inverse TQCM times the number of time steps, as a function of the experiment shots. We observe a correlation between the exponential decrease of the reconstruction error and the exponential decrease of this quantity, which saturates to a value proportional to the IPR. This is consistent with our theoretical predictions.

The relationship between IPR and information is shown in Fig.~\ref{fig:err_2q}(c), where, for a set of random initial states, we represent the IPR and the Frobenius norm of the inverse TQCM at the final time. We observe that these functions are positively correlated for small values of the TQCM, confirming the predictions of Eq. (\ref{eq:IPRvsTQCM}): to improve the performance of the learning algorithm, we have to prepare initial states with a small IPR. In particular, the best possible performance corresponds to the minimum IPR, which, for a $2$-qubit system, is $\text{IPR}_\text{min} = 1/4$.

\subsection{Random 2-body Hamiltonian}
	
Here we test the learning algorithm on a system evolving with a random Hamiltonian
\begin{equation}
H=\sum_ih_iL_i,
\end{equation}
where $h_i\in[-5,5]$ and the $L_i$ are all the two-spins interactions acting on a three-spins system, represented by tensor products of two Pauli operators and the identity operator. The Hamiltonian learning algorithm is performed with a time step $\delta t=0.01$, $N_M=1000$ measurement repetitions and a maximum number of $370$ time steps.

\begin{figure*}
    \centering
    \includegraphics[width=2 \columnwidth]{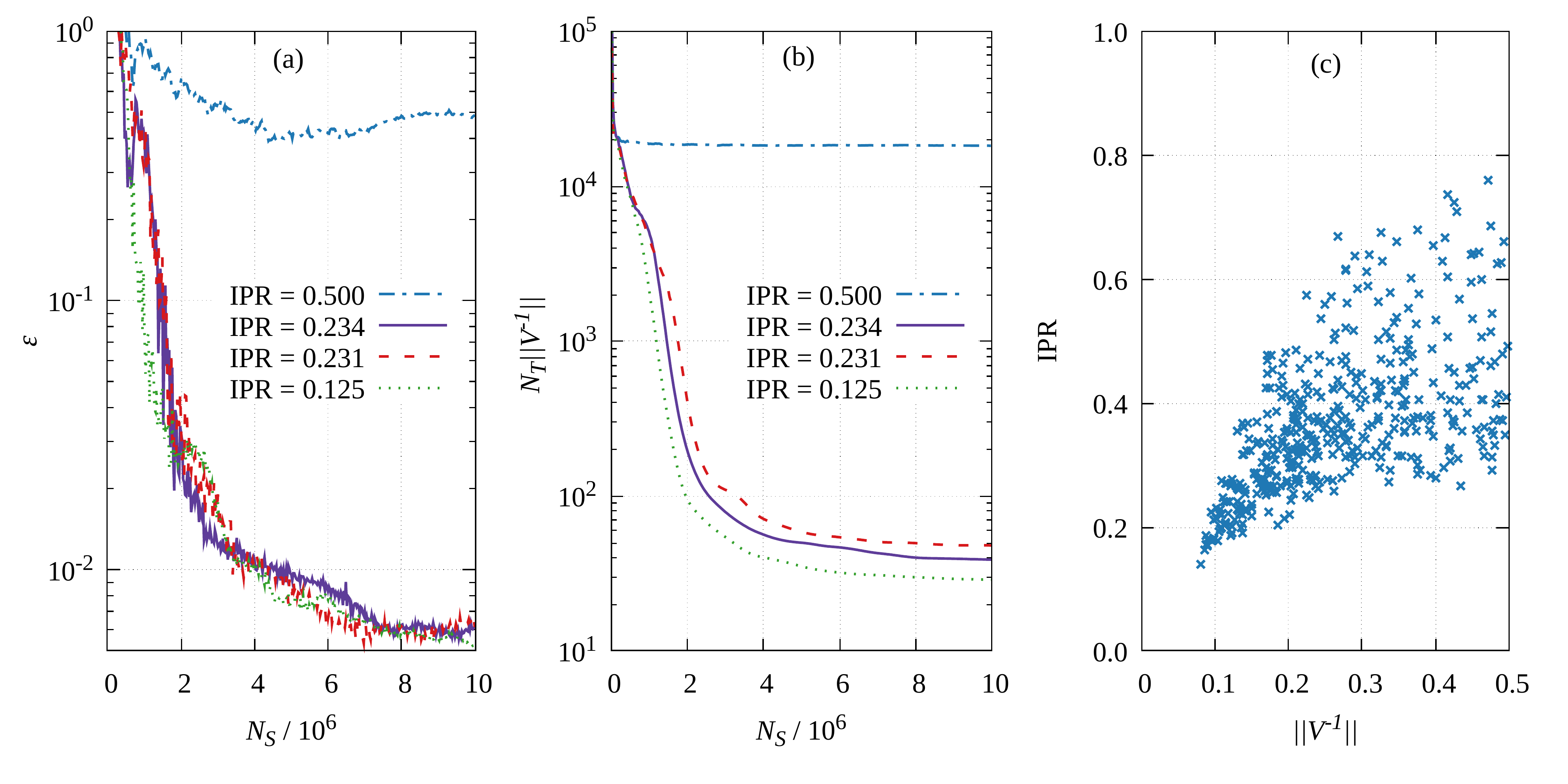}
    \caption{(a) relative reconstruction error and (b) Frobenius norm of the inverse TQCM multiplied by $N_T$, as a function of $N_S$, with $\delta t = 0.01$, $N_M=1000$ and $N_S=27\times N_M\times N_S$. The initial states are: an equal weighted superposition of two Hamiltonian eigenstates with $\text{IPR}=0.5$, the GHZ state with $\text{IPR}\approx 0.231$, $\ket{\uparrow\uparrow\uparrow}$ with $\text{IPR}\approx 0.234$, and $\ket{\psi_\text{opt}}$ with $\text{IPR}=0.125$. (c) IPR and Frobenius norm of the inverse TQCM for a collection of random states at large final time, $N_T=370$.}
    \label{fig:err_3q}
\end{figure*}

The relative error of the reconstruction and the behaviour of the TQCM are shown in Figure \ref{fig:err_3q} Panels (a) and (b). In Figure \ref{fig:err_3q} Panel (c) we show the IPR and the Frobenius norm of the inverse TQCM for a collection of random states at large final time. The validity of theoretical predictions about the optimality of low IPR states is particularly evident in Panels (b) and (c), where the statistical and systematic contributions to uncertainty are not considered.

\subsection{IBM Q FakeAthens processor}

We test the learning algorithm on a simulated quantum processor, the FakeAthens processor, using Qiskit~\cite{Qiskit}. Our approach, from state preparation to the final measurements, can be identically extended to any quantum processor. The present simulator considers a two qubits system and take into account errors in state preparation and in measurements. We remark that, due to preparation errors, the starting state is not a pure state, nevertheless our method is applicable.

We execute a time-dependent unitary gate, represented in the computational basis as
\begin{equation}
    U(t)=\left(
\begin{array}{cccc}
 \cos (4 \pi  t) & -i \sin (4 \pi  t) & 0 & 0 \\
 -i \sin (4 \pi  t) & \cos (4 \pi  t) & 0 & 0 \\
 0 & 0 & 1 & 0 \\
 0 & 0 & 0 & 1 \\
\end{array}
\right).
\end{equation}
We know that this gate is obtained through the cross-resonance mechanism illustrated in our first example, hence we look at a parent Hamiltonian that is spanned by the same operators:
\begin{align}
    \{L_i\}=\{&\id\otimes\sigma_x, \id\otimes\sigma_y, \id\otimes\sigma_x, \sigma_z\otimes\id, \sigma_z\otimes\sigma_x,\nonumber\\
    &\sigma_z\otimes\sigma_y, \sigma_z\otimes\sigma_z\}.
\end{align}

\begin{figure*}
    \centering
    \includegraphics[width=2 \columnwidth]{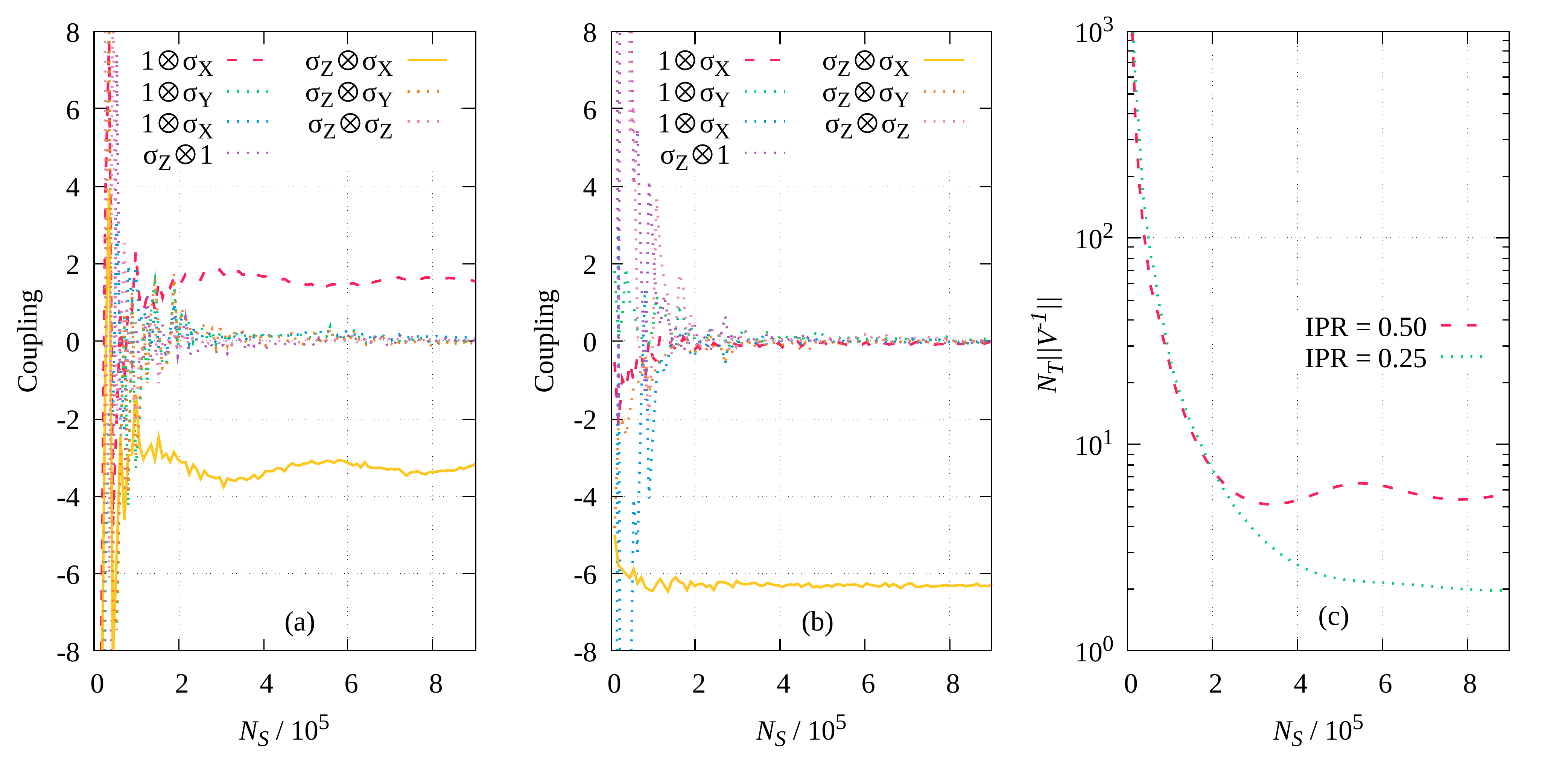}
    \caption{Panels (a) and (b): reconstructed Hamiltonian couplings at increasing number of shots, corresponding to longer observation time, for $N_M=1000$, $\delta t=0.01$ and maximum number of time steps $N_{T,\text{max}}=100$. Panel (a): initial state $\ket{\downarrow\downarrow}$, Panel (b): initial state is the Bell state. Panel (c): Frobenius norm of the inverse TQCM multiplied by $N_T$, as a function of $N_S$ for the two initial states.}
    \label{fig:IBM}
\end{figure*}

Since in this case we do not know \textit{a priori} the real system Hamiltonian, in Figure \ref{fig:IBM} Panels (a) and (b) we show, for increasing total observation time corresponding to an increasing number of shots, the Hamiltonian couplings learned with different initial states, respectively the states $\ket{\downarrow\downarrow}$ and the Bell state $(\ket{\uparrow\uparrow}+\ket{\downarrow\downarrow})/\sqrt{2}$. Looking at Panel \ref{fig:IBM}(c) we can see that a smaller uncertainty, and therefore a more reliable Hamiltonian reconstruction is obtained when the starting state is the Bell state, with smaller IPR [Panel (b)].

\section{Conclusions and outlook} 
We have introduced a Hamiltonian learning algorithm whose global accuracy  depends on the IPR of the starting state in the Hamiltonian eigenstates. The reconstruction error decays exponentially at short times and equilibrate to a value proportional to the IPR. Our results establish a direct connection between state delocalization and Hamiltonian learning. We conclude that delocalization can be exploited to drastically improve the efficiency of Hamiltonian learning algorithms. Moreover, since the TQCM can be interpreted as an information matrix on the space of states, the relationship between the time evolution of the QCM and the IPR of the state opens a new perspective on the information-geometric approach to the investigation of many-body quantum systems~\cite{qgt1,qgt2,qgt3,qgt4}: the equilibration and ergodicity of a closed quantum system~\cite{nonequilibrium} can reflect on its out-of-equilibrium geometry~\cite{qgtaway}.

Remarkably, our method is designed to reconstruct the system Hamiltonian from a generic quantum state, either pure or mixed. This feature is appealing for the application to real devices, where the preparation of the initial state can be affected by SPAM noise. In this regard, a natural extension of the methods described in this paper is the design of Lindbladian learning algorithms to deal with open quantum systems.

After the initial time transient, the errors on the learning is negligible and our results  show an excellent agreement between the true Hamiltonian used to generate the dynamics and our optimal reconstructed  Hamiltonian. The price to pay is to perform  a complete tomography of the state at different times, obtained by measuring all the observables in $\{O_\alpha\}$. The number of these observables increases exponentially in the system size, rendering our approach too demanding for large systems. This great effort in collecting expectation values is partially rewarded by an exponentially decreasing relative error [Eq.~\eqref{eq:rel_err_scaling}]. In perspective, we could perform a partial tomography and only measure the expectation values of the polynomial set of observables that optimizes the accuracy. These observables, as well as optimal initial states, could be selected through adaptive learning strategies, in analogy with Ref.~\cite{dutt2021active}, where the role of the TQCM is taken by the Fisher Information Matrix.

We thank G.~Acanfora, A.~Dutt, R.~Fazio, A.~Mezzacapo, and A.~Russomanno for useful discussions and support. This work has been funded by project code PIR01\_00011 “IBiSCo”, PON 2014-2020, for all three entities (INFN, UNINA and CNR). 

\bibliographystyle{unsrtnat}
\bibliography{refs}

\onecolumn
\appendix

	\section{Details of uncertainty estimation}\label{sec:appendix}

	In this appendix, we illustrate the details of the derivation of the bound in Eq.~(\ref{eq:rel_err_scaling}).
	
	The uncertainty on $B_i$ is
	\begin{equation}
	\delta B_i = \sqrt{\sum_{\gamma,m} \left(\frac{\partial B_i}{\partial r_\gamma(t_m)}\sigma(r_{\gamma}(t_m))\right)^2}+\sum_m\lvert \tr\left(-i[L_i,\rho(t_m)]R_m\right)\rvert
	\end{equation}
	We want to find an upper bound on this uncertainty that does not depend on the states $\{\rho(t_m)\}$.
	
	The first term in the previous equation contains the statistical uncertainty. In a spin system we can choose basis operators as normalized tensor products of Pauli operators, hence $O_\alpha^2=\id/2^{N_q}$ and $\sigma(r_\alpha(t_n))=1/\sqrt{2^{N_q}N_M}$. It follows that
	\begin{equation}
	\delta B_i = \frac{1}{\sqrt{2^{N_q}N_M}}\sqrt{\sum_{\gamma,m} \left(\frac{\partial B_i}{\partial r_\gamma(t_m)}\right)^2}+\sum_m\lvert \tr\left(-i[L_i,\rho(t_m)]R_m\right)\rvert
	\end{equation}	
	Since
	\begin{equation}
	B_i=\sum_{n\alpha\beta}\tr\left(-i[L_i,O_\alpha]O_\beta\right)r_\alpha(t_n)(r_\beta(t_{n+1})-r_\beta(t_n))/\delta t,
	\end{equation}
	we can take the derivative to obtain
	\begin{align}
	\sum_{\gamma,m} \left(\frac{\partial B_i}{\partial r_\gamma(t_m)}\right)^2&=\sum_{\gamma,m} \left(\sum_{n,\alpha\beta}\tr(-i[L_i,O_\alpha]O_\beta)\Big[\delta_{\alpha\gamma}\delta_{nm}\frac{r_\beta(t_{n+1})-r_\beta(t_{n})}{\delta t}+r_{\beta n}\delta_{\gamma \alpha}\frac{\delta_{mn}-\delta_{m, n+1}}{\delta t}\Big]\right)^2\notag\\
	&=\sum_{\gamma,m} \left(\sum_\beta\tr(-i[L_i,O_\gamma]O_\beta)\left[\frac{r_\beta(t_{m+1})-r_\beta(t_{m})}{\delta t}+\frac{r_{\beta}(t_m)-r_{\beta}(t_{m-1})}{\delta t}\right]\right)^2+o(N_T)\notag\\
	&\approx\sum_{\gamma,m}\left(\sum_\beta\tr(-i[L_i,O_\beta]O_\gamma)\left[\frac{r_\beta(t_{m+1})-r_\beta(t_{m-1})}{\delta t}\right]\right)^2.
	\end{align}
	At this point we approximate the fraction with the derivative and exploit the fact that $\{O_\alpha\}$ is an orthonormal basis:
	\begin{align}
	\sum_{\gamma,m} \left(\frac{\partial B_i}{\partial r_\gamma(t_m)}\right)^2&\approx4\sum_{\gamma,m} \left(\sum_\beta\tr(-i[L_i,O_\beta]O_\gamma)\partial_t r_\beta(t_{m})\right)^2\nonumber\\
	&=4\sum_{\gamma,m} \left(\tr(-i[L_i,\partial_t \rho(t_m)]O_\gamma)\right)^2\nonumber\\
	&=4\sum_{\gamma,m}\left(\tr\left(-i[L_i,\partial_t \rho(t_m)]\right)^2\right).
	\end{align}
	
	Replacing this estimate of the statistical uncertainty and the Taylor remainder $R_n = \frac{-[ H,[ H,\rho(t_n^*)]]}{2}$ in the total uncertainty $\delta B_i$, we obtain
	\begin{equation}
	\delta B_i \approx \frac{2}{\sqrt{2^{N_q}N_M}}\sqrt{\sum_{m} \left(\tr\left(-i[L_i,\partial_t \rho(t_m)]\right)^2\right)}+\frac{\delta t}{2}\sum_m\lvert 	\tr\left[\left(-i[L_i,\rho(t_m)]\right)\left(-[ H,[ H,\rho(t_m^*)]]\right)\right]\rvert,
	\end{equation}
	where $ H$ is the system Hamiltonian.

	Considering the Cauchy-Schwarz inequality $\lvert \tr(AB)\rvert \leq\sqrt{\tr(AA)\tr(BB)}$, this last estimate can be bounded as
	\begin{equation}
	\delta B_i \leq \frac{2}{\sqrt{2^{N_q}N_M}}\sqrt{\sum_{m} \left(\tr\left(-i[L_i,-i[ H, \rho(t_m)]]\right)^2\right)}+\frac{\delta t}{2}\sum_m\sqrt{\tr\left(\left(i[L_i,\rho(t_m)]\right)^2\right)\tr\left([ H,[ H,\rho(t_m^*)]]^2\right)},
	\end{equation}
	where we have taken into account that $\partial_t\rho(t_m)=-i[H,\rho(t_m)]$.
	
	Now we need to understand how the commutator with an Hermitian operators changes the Frobenius norm of a given operator. In particular, given a Hermitian operator $A$ with spectral decomposition $A=\sum_i a_i \ket{i}\bra{i}$, we define $a_\text{min}=\min(a_i)$, $a_\text{max}=\max(a_i)$, and $ A_\delta = A-a_\text{min}\id$. Hence we can write
	\begin{align}
	\tr[(-i[A,X])^2]&=\tr[(-i[A_\delta,X])^2]\notag\\
	&=2[\tr(XXA_\delta A_\delta)-\tr(A_\delta XA_\delta X)]\notag\\
	&=2\sum_{ij}(a_i-a_\text{min})(a_i-a_j)\lvert \bra{i}X\ket{j}\rvert ^2\notag\\
	&\leq 2 (a_\text{max}-a_\text{min})^2\sum_{i j}\lvert \bra{i}X\ket{j}\rvert ^2\notag\\
	&= 2 \lVert A_\delta\rVert _\text{op}^2\tr(X^2),
	\end{align}
	where $\lVert A_\delta \rVert _\text{op}$ is the operator norm of $A_\delta$.
	
	Replacing this bound in our estimate of the uncertainty $\delta B_i$ and considering that the purity of the state $\rho(t)$ is $\tr(\rho(t)^2)\leq 1$ for each value of $t$, we obtain
	\begin{align}
	\delta B_i &\leq \frac{2 \lVert L_i\rVert _\text{op}}{\sqrt{2^{N_q}N_T}}\sqrt{\sum_{m}2 \left(\tr\left(-i[ H, \rho(t_m)]\right)^2\right)}\nonumber+\lVert L_{i,\delta}\rVert _\text{op}\lVert  H_\delta\rVert _\text{op}\sum_m\sqrt{\tr\left(\left(-i[ H,\rho(t_m^*)]\delta t\right)^2\right)}\nonumber\\
	&\leq \frac{4 \lVert  H_\delta\rVert _\text{op} \lVert L_{i,\delta}\rVert _\text{op}}{\sqrt{2^{N_q}N_M}}\sqrt{N_T}+\delta t\lVert L_{i,\delta}\rVert _\text{op}\lVert  H_\delta\rVert _\text{op}^2N_T,
	\end{align}
	where $\lVert L_{i,\delta}\rVert _\text{op}$ and $\lVert  H_{\delta}\rVert _\text{op}$ are the difference between the maximum and the minimum eigenvalues of $ H$ and $L_i$, respectively.
	
	For a traceless operator $A=\sum_i a_i \ket{i}\bra{i}$ we have that the maximum eigenvalue is positive and the minimum one is negative, therefore
	\begin{equation}
	\lVert A_\delta \rVert _\text{op}=a_\text{max}-a_\text{min}=\lvert a_\text{max}\rvert+\lvert a_\text{min}\rvert\leq 2\text{max}(\lvert a_i\rvert)=2\lVert A\rVert_\text{op}.
	\end{equation}
	Hence, since both $ H$ and $L_i$ are traceless and since we can choose without loss of generality $\lVert L_i\rVert_\text{op} = \lVert L\rVert_\text{op} \forall i$, the last inequality becomes
	\begin{equation}
	\delta B_i \leq \frac{16 \lVert  H\rVert _\text{op} \lVert L\rVert _\text{op}}{\sqrt{2^{N_q}N_M}}\sqrt{N_T}+4\delta t\lVert L\rVert _\text{op}\lVert  H\rVert _\text{op}^2N_T.
	\end{equation}
	
	An analogous bound on the uncertainty about the Hamiltonian couplings can be calculated propagating the uncertainty about $B_i$. When $\lVert V^{-1}\rVert \ll 1$ we obtain
	\begin{equation}
	\delta h_i = \sum_{j}(V^{-1})_{ij}\delta B_j.
	\end{equation}
	and therefore
	\begin{equation}
	\lVert \vec \delta h\rVert  \leq \sqrt{\tr \left(V^{-2}\right)} \lVert L\rVert _\text{op}\left(16 \lVert  H \rVert _\text{op}\sqrt{\frac{N_T}{2^{N_q}N_M}} +4N_T\lVert  H\rVert_\text{op}^2\delta t\right).
	\end{equation}
	When $\lVert V^{-1}\rVert \ll 1$ does not hold, this estimate for the uncertainty fails, but this is the case in which we have an uncertainty that is so large that finding the exact Hamiltonian is impossible.

	Now, we want to derive a bound on the relative uncertainty on the couplings.  Taking into account the triangular inequality and the relationship between $p$-norms, and defining $l$ as the number of Hamiltonian couplings, we can write
	\begin{equation}
	\lVert  H\rVert_\text{op}=\lVert \sum_i  h_i L_i \lVert_\text{op}\leq \sum_i \lvert h_i\rvert\lVert  L_i \rVert_\text{op} =  \lVert  L \rVert_\text{op}\sum_i \lvert h_i \rvert\leq \sqrt{l}\lVert  L \rVert_\text{op}\lVert\vec{ h}\rVert,
	\end{equation}
	from which we finally obtain
	\begin{equation}
	\frac{\lVert \vec \delta h\rVert }{\lVert\vec { h}\rVert} \leq \sqrt{l\tr \left(V^{-2}\right)} \lVert L\rVert _\text{op}^2\left(16 \sqrt{\frac{N_T}{2^{N_q}N_M}} +4N_T\lVert  H\rVert _\text{op}\delta t\right).
	\end{equation}

\end{document}